\documentclass[fleqn,11pt]{wlscirep}
\usepackage[utf8]{inputenc}
\usepackage[T1]{fontenc}
\usepackage{bm}
\title{Suppression of axionic charge density wave and onset of superconductivity in the chiral Weyl semimetal Ta$_2$Se$_8$I}

\author[1]{Qing-Ge Mu}
\author[1,2]{Dennis Nenno}
\author[1,3]{Yan-Peng Qi}
\author[1]{Feng-Ren Fan}
\author[3]{Cuiying Pei}
\author[1]{Moaz ElGhazali}
\author[1,4]{Johannes Gooth}
\author[1]{Claudia Felser}
\author[2,*]{Prineha Narang}
\author[1,**]{Sergey Medvedev}
\affil[1]{Max Planck Institute for Chemical Physics of Solids, 01187 Dresden, Germany.}
\affil[2]{John A. Paulson School of Engineering and Applied Sciences,
Harvard University, Cambridge, Massachusetts 02138, USA.}
\affil[3]{School of Physical Science and Technology, ShanghaiTech University, Shanghai 201210, China.}
\affil[4]{Institut fur Festkorper- und Materialphysik, Technische Universitat Dresden, 01069 Dresden, Germany.}
\affil[*]{Email: prineha@seas.harvard.edu}
\affil[**]{Email: Sergiy.Medvediev@cpfs.mpg.de}

\newcommand{\TSI}[0]{Ta$_2$Se$_8$I}
\newcommand{\TCDW}[0]{$T_{\mathrm{CDW}}$}
\newcommand{\Tc}[0]{$T_c$}

\begin{abstract}
A Weyl semimetal with strong electron-phonon interaction can show axionic coupling in its insulator state at low temperatures, owing to the formation of a charge density wave (CDW). Such a CDW emerges in the linear chain compound Weyl semimetal \TSI~ below 263 K, resulting in the appearance of the dynamical condensed-matter axion quasiparticle. In this study, we demonstrate that the interchain coupling in \TSI~ can be varied to suppress the CDW formation with pressure, while retaining the Weyl semimetal phase at high temperatures. Above 17 GPa, the Weyl semimetal phase does not survive and we induce superconductivity, due to the amorphization of the iodine sub-lattice. Structurally, the one-dimensional Ta-Se-chains remain intact and provide a superconducting channel in one dimension. We highlight that our results show a near-complete suppression of the gap induced by the axionic charge-density wave at pressures inaccessible to previous studies. Including this CDW phase, our experiments and theoretical predictions and analysis reveal the complete topological phase diagram of \TSI~ and its relationship to the nearby superconducting state. The results demonstrate \TSI~ to be a distinctively versatile platform for exploring correlated topological states.
\end{abstract}
\begin{document}

\flushbottom
\maketitle

\thispagestyle{empty}

Topological materials are typically well-described by the band theory of non-interacting electrons.~\cite{qi_topological_2008,RevModPhys.82.3045,RevModPhys.90.015001} However, for materials with strong electron-phonon interactions, this description is no longer valid.~\cite{Yang2011,Laubach2016,wang_chiral_2013,wieder2020dynamical} Instead, their topological ground state merely emerges from minimizing the energy of the strongly coupled electronphonon system. 
A Weyl semimetal with strong electron-phonon interactions, for example, can show axionic coupling in its insulator state at low temperatures, owing to the formation of a charge density wave (CDW).~\cite{gooth_axionic_2019,wang_chiral_2013,wieder2020dynamical} In their parent state, Weyl semimetals are materials in which low-energy electronic quasiparticles behave as chiral relativistic fermions without rest mass, known as Weyl fermions.~\cite{RevModPhys.90.015001} Weyl fermions exist at isolated crossing points of the conduction and valence bands -- the Weyl nodes -- and their energy can be approximated with a linear dispersion relation. By turning on strong electron-phonon interactions, a CDW can emerge that links two Weyl nodes of different chiralities and gaps the Weyl crossing points. 
A charge density wave (CDW) is an ordered quantum fluid of electrons that forms a standing wave pattern along the atomic chains. It is characterized by a gap in the single-particle excitation spectrum and by a gapless electrical-current-carrying collective mode, the phason.\cite{anderson2018basic,monceau_electronic_2012}
In Weyl semimetals, the phason of a charge density wave state is an axion, which couples to an electromagnetic field in the topological $\theta E{\cdot}B$ term.~\cite{wang_chiral_2013,axion_review_2020} Recently, signatures of such an axionic CDW were found in the linear-chain-compound Weyl semimetal, resulting in the observation of the dynamical condensed-matter axion quasiparticle.~\cite{gooth_axionic_2019}
\TSI~ is a quasi-one-dimensional material with a body-centered tetragonal lattice (see Fig.~\ref{fig3}(c)). The Ta atoms are surrounded by Se$_4$-rectangles and form chains aligned along the c-axis. These chains are separated by I-ions. Upon cooling down below \TCDW~ = 263 K, \TSI~ undergoes a CDW transition at ambient pressure.\cite{fujishita-a,forro-a,forro-b,lorenzo1998a,nemeth2001a,tournier-colletta2013a} The wave vector of the CDW in \TSI~ is incommensurate with the lattice, and the phason is generally pinned to impurities. Therefore, only upon applying a certain threshold electric field (above which the electric force overcomes the pinning forces) the phason is depinned and free to slide over the lattice, thereby contributing to the electrical conduction. The resulting conduction behavior is strongly nonlinear and is a characteristic feature of the CDW state.
\TSI~ is so far the only material in which a combination of charge-density wave and chiral anomaly have been demonstrated, and experimental conditions sensitively influence the observability of this phase.~\cite{gooth_axionic_2019,cohn2020magnetoresistance} In order to map out the region of existence for this axionic charge density wave, we combine magneto-transport measurements, Raman spectroscopy and \emph{ab initio} calculations to assess its phase diagram. For pressures below the superconducting transition, we show a near-complete suppression of the CDW that has remained elusive in prior work. Above the critical pressure at which we observe superconductivity, our results highlight that the TaSe$_4$ building blocks of the quasi-one dimensional chains survive, while the rest of the structure becomes amorphous. The present study accesses the phase diagram beyond what is known about this unique compound and paves the way toward a more microscopic understanding of its dynamics.
The samples used in this study are the same as used in Ref.~\citeonline{gooth_axionic_2019}.

\begin{figure*}[t]
    \centering
    \includegraphics[width=1.0\textwidth]{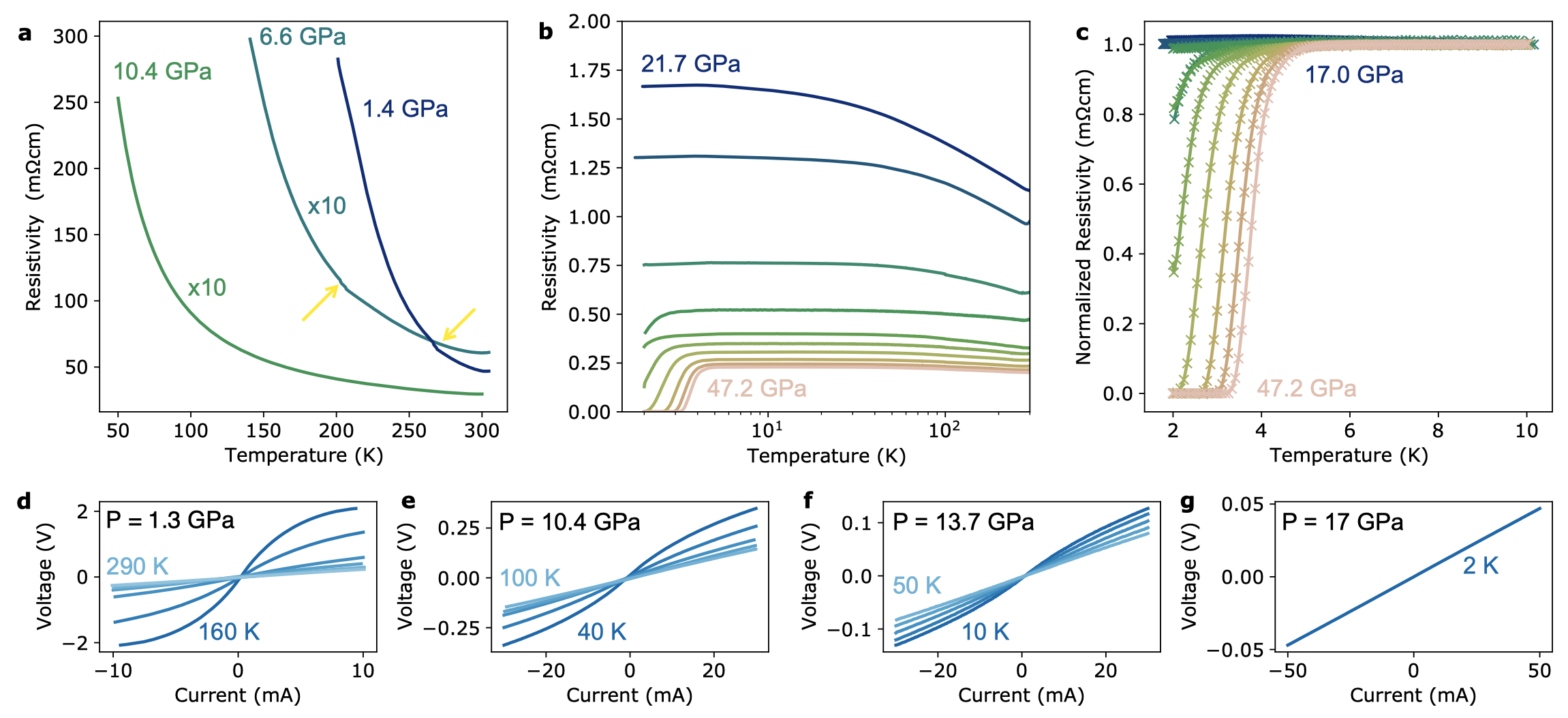}
    \caption{Electrical transport in under pressure. (a) Electrical resistivity versus temperature of \TSI~ due to the formation of a charge density wave (\TCDW~ indicated with yellow arrows) is visible at low pressures. (b) The V-I characteristic switches from linear behavior (Ohmic) at high temperatures to nonlinear behavior by cooling below the T$_{\mathrm{CDW}}$, which is a signature of sliding CDW transport. At higher pressures, the kink in the resistivity curves is smeared out (resistivity curve at 10.4 GPa in (a)). The V-I characteristics, however, are demonstrating the switch to nonlinear behavior at high pressures for progressively lower temperatures (d-f). Finally, the V-I characteristic remains linear up to the lowest temperature as pressure increased to 17 GPa (g), indicating the complete suppression of CDW-formation. A drop of resistivity is observed at low temperature in resistivity curves as pressure increased further beyond the suppression of CDW formation (b), which suggests the onset of a superconducting transition. The superconducting state is clearly observed at higher pressures, with the resistivity dropping abruptly to zero (c) while the resistivity in normal state demonstrates rather semimetallic character (b). The \Tc~keeps increasing monotonically with pressure further increase, and reaches 4.5 K at the highest pressure 47.2 GPa.}
    \label{fig1}
\end{figure*}{}

We first map the transition temperature of the charge density wave under pressure using resistivity measurements. Figure~\ref{fig1}(a) shows the single-particle electrical resistivity $\rho$ of \TSI~ as a function of temperature $T$ at various pressures $P$ up to 10.4 GPa. At low $P$ (1.4 GPa and 6.6 GPa), each $\rho(T)$ exhibits a faint kink, i.e.~a peak in its derivative (see SI for details), at particular temperatures \TCDW, indicative of the formation of the CDW (see SI for details).\cite{gooth_axionic_2019} The CDW transition temperature of our samples at ambient $P$ is \TCDW = 263 K,\cite{gooth_axionic_2019} in good agreement with existing literature. With increasing pressure, we observe that the \TCDW~first increases slightly until 1.4 GPa to 267 K, but is then suppressed down to 204 K at 6.6 GPa (Fig. 1a). This observed increase at low pressures is in agreement with previous reports.\cite{nunezregueiro1993a,forro-b} Consistent with a sliding phason mode, we observe nonlinearity in the VI curves below \TCDW~ at high bias currents, that is, high voltages. For $P$ > 10 GPa, the signatures of the CDW transition in $\rho(T)$ is blurred out, but the crossover from linear V-I characteristics at high T to non-linear V-I characteristics at low T remains observable. Hence, in order to obtain the transition temperature for all pressures investigated, we use the onset temperature $T_{on}$ of the non-linear behavior in the V-I characteristic as a measure for \TCDW~ (see SI for details). This procedure is justified by the experiments below 6.6 GPa, for which the independently determined \TCDW~ and $T_{on}$ match and allows us to go beyond the territory of phase diagrams in previous studies and to observe the near-closing of the gap induced by the charge-density wave.~\cite{monceau_electronic_2012} 
Consistently, we find that \TCDW~ (= $T_{on}$) for pressures above 6.6 GPa decreases with increasing $P$. Finally, the V-I characteristic becomes linear at the lowest temperature 2 K investigated as the pressure is increased to 17 GPa (Supplementary materials), indicating the complete suppression of CDW-formation. 
Hall measurements above \TCDW~ show that the carrier concentration increases with increasing $P$, indicating a enhancement of the Fermi-surface area when $P$ is enhanced. This observation suggest that the Weyl points move away from the Fermi level for increasing pressure and potential contributions due to interband coupling.
To gain more insights into the CDW state below \TCDW, we fit the single particle $\rho(T)$ in this temperature regime with a thermal activation law,
\begin{equation}
    \rho(T) \propto \exp\left(\frac{\Delta E}{2k_{\mathrm{B}}T}\right) \,,
    \label{eq:rhoT}
\end{equation}
with the energy gap $\Delta E$ and the Boltzmann constant $k_{\mathrm{B}}T$ (Supplementary materials). The estimated single-particle gap at low temperature decreases with increasing pressure from $\Delta E$ = (259 $\pm$ 14) meV at ambient pressure to $\Delta E$ = (18.3 $\pm$ 0.03) meV at 10.4 GPa, again, consistent with the decreasing \TCDW.

\begin{figure*}[t]
    \centering
    \includegraphics[width=0.8\textwidth]{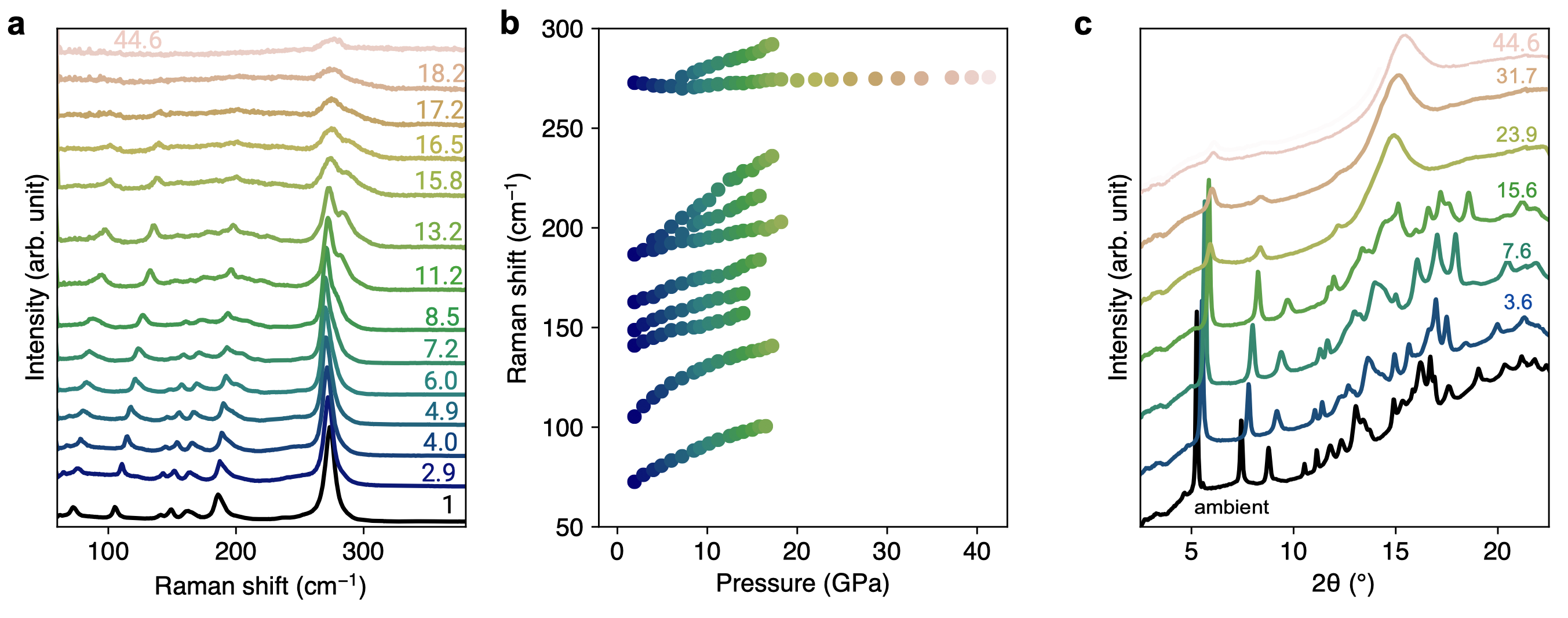}
    \caption{Pressure effect on structure of \TSI. Upon applying pressure, all observed Raman peaks (a) except the one located at 274 cm$^{-1}$ shift to higher frequencies (b) demonstrating normal stiffening behavior upon compression. Disappearance of the majority of Raman peaks at pressures above 17 GPa (a) indicates pressure-induced amorphisation of \TSI. \textit{In situ} synchrotron XRD patterns of \TSI~ (c) demonstrate the pressure-induced amorphisation as pressure approaching $\sim$20 GPa. The strongest Raman peak (located at $\sim$274 cm$^{-1}$) corresponding to the intrachain Se-Se stretching vibrations can be observed up to the highest in these experiments pressures (a, b) indicating the preservation of the TaSe$_4$-units as a building blocks of local order in the high-pressure amorphous phase of \TSI.}
    \label{fig2}
\end{figure*}{}

Remarkably, above $P$ = 20 GPa -- the pressure where the CDW is suppressed --- we observe a drop of $\rho(T)$ at low temperatures  (Fig.~\ref{fig1}(b)), indicating the onset of a superconducting transition. The superconducting state is clearly observed at higher pressures, with the single-particle resistivity dropping abruptly to zero when reaching the transition temperature \Tc~ (Fig.~\ref{fig1}(c)). With increasing pressure, \Tc~ keeps increasing monotonically and is 4.5 K at our maximum pressure of 47.2 GPa. See supplementary materials for details and further study of the superconducting state. The state above \Tc~ remains semimetallic. Observations of pressure-induced superconductivity in \TSI~ are in accordance with the results obtained on a number of quasi-1D transition metal chalcogenides in which suppression of the CDW-state leads to the emergence of superconductivity.\cite{yomo2005a,monteverde2013a,nunezregueiro1993a,yang2018a,yasuzuka2005a,gu2018a} We find that the superconductivity in \TSI~ emerges only after the CDW-state is completely suppressed while the superconductivity and CDW transition coexist in related compounds, e.g. ZrTe$_3$, orthorhombic-TaS$_3$, NbSe$_3$.\cite{yomo2005a,monteverde2013a,yasuzuka2005a} Such coexistence of superconductivity and CDW-transition in o-TaS$_3$ is proposed to be related to the existence of a quantum critical point  a point at which the CDW-transition temperature is driven to zero by pressure.\cite{monteverde2013a} Quantum fluctuations diverge at this point and might induce superconductivity with dome-shaped pressure dependence of \Tc~ in which maximum \Tc~ appearing around this quantum critical point. In \TSI~, however, pressure drives the system further away from the quantum critical point while the \Tc~ continuously increases. These results show that maxima in \Tc~ value in such systems might be obtained by going farther away from CDW-instability, and that a dome-like shape for pressure dependence of \Tc~ can be obtained without relation to CDW order.\cite{tissen2013a}

\begin{figure}[t]
    \centering
    \includegraphics[width=\textwidth]{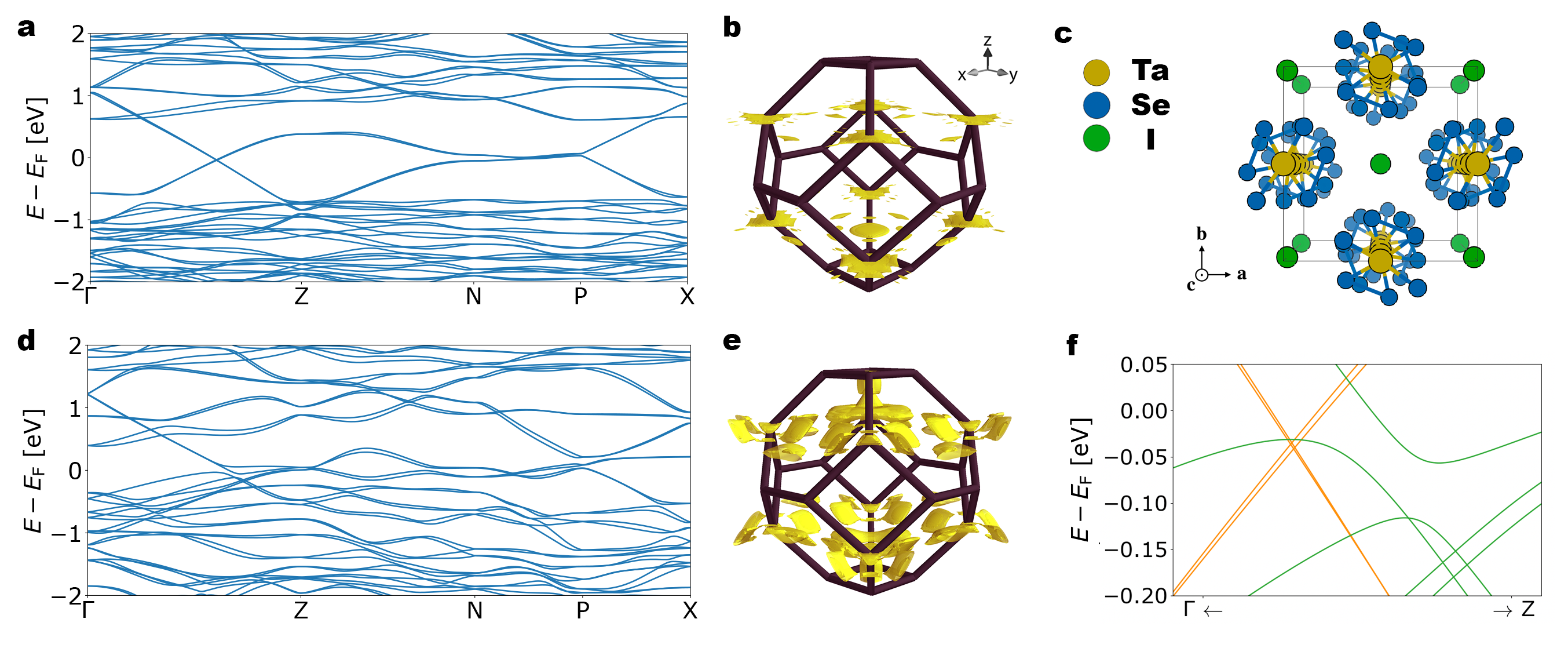}
    \caption{\emph{Ab initio} electronic band structure in the high-symmetry phase of \TSI~ at 7.6 GPa (a) and 15.6 GPa (d) incorporating the experimental lattice parameters. While the Fermi surface at 7.6 GPa (b) does not yet deviate drastically from quasi-one dimensional behavior, nor from the result at ambient pressure, bands with contributions from the Iodine atoms add to the three-dimensionality at 15.6 GPa (e). The conventional cell of \TSI~ displays Ta-Se chains spaced by iodine atoms (c). Increasing pressure leads to the Weyl point manifold on the $\Gamma \to Z$ axis moving away from the Fermi level at 7.6 GPa (orange lines, f) and 15.6 GPa (green lines, f).}
    \label{fig3}
\end{figure}{}

To gain a deeper understanding of the different states in the pressure phase diagram in \TSI, we performed pressure-dependent Raman spectroscopy studies. The resulting spectra at selected pressures are shown in Figure \ref{fig2}(a) and are consistent with previous reports.\cite{ikari1985a,sekine-a,zwick1985a,An_AdvMat_2020} Up to ~18 GPa, the crystal structure of \TSI ~becomes increasingly asymmetric with higher pressure but remains stable. In particular, all occurring peaks persist at their position or shift to higher frequencies upon increasing pressure, but remain observable (Fig.~\ref{fig2}(b)). At pressures around 7 GPa, the Raman spectra demonstrate splitting of selected modes (Fig.~\ref{fig2}), which can indicate a possible structural phase transition. However, the peak splitting is observed only for selected types of modes, and there are no sudden changes in spectra upon pressure increase. Therefore, the possibility of a major structural transition can be ruled out since such a transition is usually associated with drastic changes of the Raman spectra due to the deformation of the Brillouin zone. On one hand, considering that the peaks consist of multiple degenerate modes (Supplementary materials), the peak splitting can also be ascribed to different pressure coefficients of the respective modes. On the other hand, this splitting might be indicative for the distortion of the Ta-Ta distances within the chains (which are uniform at ambient pressure) resulting in the continuous distortion of the initially slightly distorted rectangular Ta-Se antiprisms, similar to the case of (NbSe$_4$)$_3$I with slightly distorted Nb-Nb distances in which corresponding Raman peaks are split.\cite{ikari1985a} It is worth noting the remarkable similarity of the unusual non-monotonic behavior of the frequencies from the strongest Raman peak in \TSI~ assigned to the intrachain Se-Se stretching vibrations (Fig.~\ref{fig2}) and the corresponding Raman-peak in Q1D o-TaS$_3$.\cite{wu2017a} This similarity is indicative for the generality in the structural response to external pressure in Q1D transition metal chalcogenides, which remains unexplored so far. 
Pressure-dependent x-ray diffraction studies (Fig.~\ref{fig2}(c)) further demonstrate the stability of the \TSI~ structure up to ~20 GPa, as no additional peaks or splittings are observed in the diffraction patterns. 
For pressures above 20 GPa, however, only the strongest high-frequency peak persists in the experimental spectra, indicating diminishing crystal order at least in part of the \TSI~ sample. This pressure-induced amorphization is observed also by the evolution of x-ray diffraction patterns in which disappearance of the sample Bragg peaks and the formation of broad amorphous halo patterns are observed at pressures above 20 GPa (Fig.~\ref{fig2}(c)). The persistent, isolated Raman peak (located at $\approx$ 274 cm$^{-1}$) corresponds to the intrachain Se-Se stretching vibrations and can be observed up to the highest pressures (above 40 GPa), indicating the preservation of the TaSe$_4$-units as building blocks of local order in the high-pressure amorphous phase of \TSI.
At sufficiently high pressures (beyond $\sim$ 18.2 GPa), only the strongest high-frequency peak persists in the experimental spectra. The strong background due to the luminescence of stressed diamond anvils screens most likely all weaker peaks. The decomposition of the \TSI~ crystal structure might be excluded, as no Raman modes that could not be assigned to the parent material were observed. Thus, the persistence of the strongest peak in the spectra and its monotonic behavior over frequency indicates that the crystal structure of \TSI~ remains intact up to the highest pressure.~\cite{An_AdvMat_2020} However, partial loss of crystallinity (or pressure-induced amorphization) cannot be excluded. Structural studies of \TSI~ under such high pressures would be of importance for further understanding of the properties of this material.

\begin{figure}[t]
    \centering    
    \includegraphics[width=0.5\textwidth]{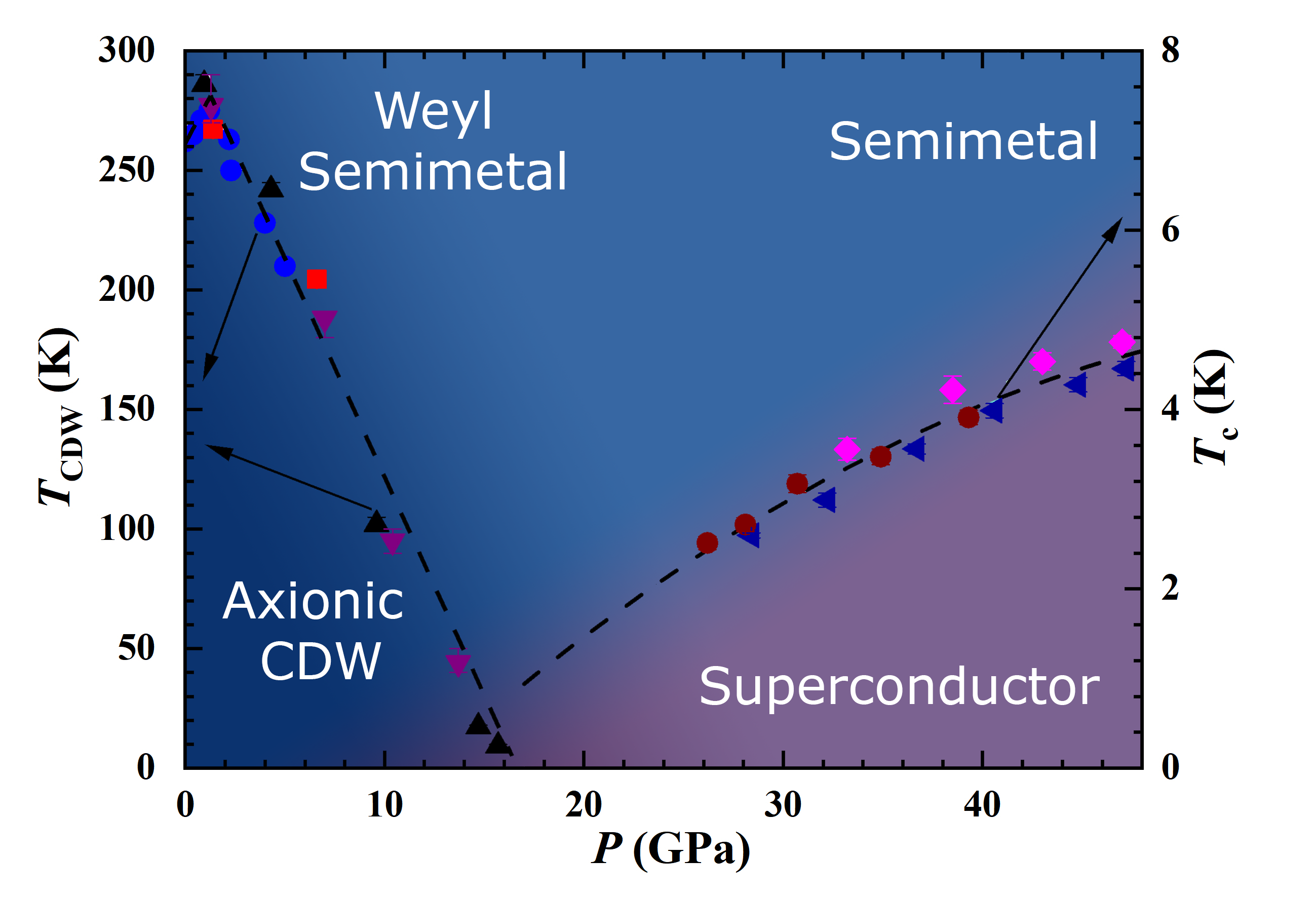}
    \caption{The phase diagram displaying the pressure effects on CDW transition and superconductivity with pressure up to 47.2 GPa on several samples. The solid blue circles are the CDW transition temperatures reported by Ref.~\citenum{nunezregueiro1993a,forro-b}. The red squares are extracted from resistivity measurements, and both purple and black triangles are from the V-I characteristics. For superconductivity, different symbols represent the onset \Tc~ from different samples. Error bars of \TCDW~ from V-I curves are based on the temperature step. Error bars of \TCDW~from resistivity and superconducting \Tc~come from average of multiple readings. }
    \label{fig4}
\end{figure}{}

Further insights in the semimetal and CDW state for pressures below the superconducting transition emerge from our \emph{ab initio} calculations. Our results for the electronic bandstructure for the high-temperature phase of \TSI~  are summarized for various pressures in Fig.~\ref{fig3}. Consistent with the conclusion from the Hall measurements, compressing the unit cell leads to a shift of the Ta d$_{3z^2-r^2}$ towards lower energies, which moves the Weyl points away from the Fermi level, reducing the net chiral charge and increasing the density-of-states at the Fermi level.~\cite{shi_charge-density-wave_2019} 
At the same time, the interchain coupling increases up until I $p$-bands hybridize at the Fermi level above 15 GPa. At this point, we expect the CDW mechanism outlined in Refs.~\citenum{shi_charge-density-wave_2019,zhang_first-principles_2020} to break down, as the charge density wave vector is incommensurate with the Fermi surface nesting, i.e. $2k_{\mathrm{F}}\neq q_{\parallel}^{\mathrm{CDW}}$, and consequently, the axionic magneto-response observed in the gapped phase should vanish.~\cite{gooth_axionic_2019} We additionally calculate the phonon dispersions in the low-temperature regime (Supplementary materials) that reveal the structural instability that leads to the formation of the CDW below 16 GPa, which was observed for ambient pressure in Ref.~\citenum{zhang_first-principles_2020,Yin_PRM_2020}.

The complete phase diagram, created with newly discovered physics in \TSI~ as well as our calculations and analysis is shown in Fig.~\ref{fig4}.~\cite{gooth_axionic_2019,an2020long} At ambient pressure, \TSI~ is a Weyl semimetal that exhibits a phase transition to an axionic CDW state below \TCDW = 263 K.~\cite{gooth_axionic_2019} With increasing pressure, our data shows that the transition temperature and consquently, the gap size, is successively reduced. At 18.2 GPa, the CDW state is fully suppressed, the Q1D behavior vanishes, and a partial amorphization accompanied with a superconducting transition emerges. The critical temperature continuously increases up to 4.5 K at the highest pressure applied (47.2 GPa). Looking ahead, structural studies would deepen the understanding of the properties of \TSI~ and similar yet-to-be-observed axionic materials under such high pressures.

\bibliography{references.bib}

\noindent\textbf{Acknowledgments}\\
This work is supported by the DARPA DSO under the Driven Nonequilibrium Quantum Systems (DRINQS) program, grant number D18AC00014. 
D.~M.~N. is partially supported by the Department of Energy `Photonics at Thermodynamic Limits Energy Frontier Research Center under grant number DE-SC0019140.
Y.Q. acknowledges the support from the National Key Research and Development Program of China (Grant No. 2018YFA0704300) and the National Natural Science Foundation of China (Grant No. U1932217 and 11974246). We also thank the staff from beam line BL15U1 at the Shanghai Synchrotron Radiation Facility for assistance during data collection
P.~N. is a Moore Inventor Fellow through Grant GBMF8048 from the Gordon and Betty Moore Foundation.
This research used resources of the National Energy Research Scientific Computing Center (NERSC), a U.S. Department of Energy Office of Science User Facility operated under Contract No. DE-AC02-05CH11231.\\

\end{document}